\documentclass[aps,prl,reprint,twocolumn]{revtex4-2}

\usepackage{amsmath, amssymb, amsfonts}

\usepackage[usenames,dvipsnames]{color}
\usepackage[colorlinks=true,linkcolor=Maroon,citecolor=Maroon,urlcolor=Maroon]{hyperref}
\usepackage{graphicx}
\usepackage{xcolor}

\makeatletter
\def\ps@pprintTitle{%
 \let\@oddhead\@empty 
 \let\@evenhead\@empty
 \def\@oddfoot{}%
 \let\@evenfoot\@oddfoot}
 \makeatother

\newcommand{\x}{\mathbf{x}}
\newcommand{\z}{\mathbf{z}}

\newcommand{\A}{\mathbf{A}}
\newcommand{\PP}{\mathbf{P}}

\newcommand{\I}{\mathbf{I}}

\newcommand{\E}{\mathbf{E}}
\newcommand{\C}{\mathbf{C}}
\newcommand{\B}{\mathbf{B}}
\newcommand{\Ss}{\mathbf{S}}
\newcommand{\G}{\mathbf{G}}

\newcommand{\rr}{\mathbf{\hat{r}}}
\newcommand{\lv}{\mathbf{\hat{l}}}

\newcommand{\pp}{\mathbf{\hat{p}}}

\newcommand{\et}{\boldsymbol{\eta}}
\newcommand{\xii}{\boldsymbol{\xi}}

\newcommand{\Sig}{\boldsymbol{\Sigma}}
\newcommand{\Gam}{\boldsymbol{\Gamma}}

\newcommand{\diag}{\mathrm{diag}}
\newcommand{\mw}[1]{\left\langle #1 \right\rangle}

\begin{document}

\title{Pseudo-Coherence and Stochastic Synchronization:\\
A Non-Normal Route to Collective Dynamics without Oscillators}

\author{V. Troude}
\author{D. Sornette}
\affiliation{Institute of Risk Analysis, Prediction and Management (Risks-X),
    Academy for Advanced Interdisciplinary Sciences,
    Southern University of Science and Technology, Shenzhen, China
}

\begin{abstract}
Collective temporal organization in complex systems is commonly attributed to synchronization, resonance, or proximity to dynamical instabilities. 
Here we identify a distinct mechanism by which coherent, synchronization-like behavior can emerge in stochastic systems that are linearly stable and contain no intrinsic oscillators. 
The mechanism arises from non-normal pseudospectral amplification and leads to what we term \emph{pseudo-coherence}: an intermittent form of collective organization characterized by transient phase alignment, broken time-reversal symmetry, positive entropy production, and drifting spectral peaks.
Using a minimal overdamped stochastic model, we show that increasing non-normality drives a sharp pseudo-critical transition. 
Beyond a well-defined threshold, fluctuations concentrate along a dominant reaction mode, generating intermittent growth of Kuramoto-like order parameters and irreversible probability currents without eigenvalue crossings or Hopf bifurcations. 
Analytically, we demonstrate that pseudo-critical non-normal dynamics reshapes the imaginary pseudospectrum, amplifying slow fluctuations and producing coherent frequency bands under finite-time observation.
These results identify pseudo-coherence as a new route to collective temporal organization in non-equilibrium systems, suggesting that apparent rhythms and synchronization in natural systems may arise from non-normal stochastic amplification rather than intrinsic oscillators.

\textbf{Significance}: \emph{Pseudo-coherence uncovers a new organizing principle for collective dynamics, whereby non-normal stochastic amplification generates transient coherence, irreversibility, and emergent spectral structure in systems that remain linearly stable and oscillator-free. 
This mechanism provides an alternative explanation for rhythmic activity and synchronization-like patterns widely observed in biological, ecological, and physical systems where no intrinsic oscillators are known to exist. Because it arises purely from linear geometry and stochastic forcing, pseudo-coherence is expected to be widespread in high-dimensional systems with non-normal interactions. Such interactions are themselves generic, arising naturally in systems with asymmetric couplings and approximately hierarchical or structured interaction architectures. Our results therefore suggest that apparent oscillations, clustering, and arrows of time in empirical data may often reflect non-normal amplification rather than genuine limit cycles or phase-locking dynamics.}\end{abstract}

\maketitle

\section{Introduction}

Collective temporal organization is a hallmark of complex systems.
From synchronized oscillations in biological populations to quasi-periodic variability in climate and geophysical records,
coherent temporal patterns are often interpreted as signatures of intrinsic oscillators, resonant modes, or proximity to critical instabilities \cite{strogatz2003sync,pikovsky2003synchronization,kuramoto1984chemical}.
Classical theories of synchronization and critical phenomena have provided powerful tools to analyze such behavior,
yet they rest on a common premise: that coherence reflects either nonlinear phase locking between oscillatory units
or the emergence of unstable or marginally stable modes \cite{strogatz2003sync,pikovsky2003synchronization,stanley1999scaling}.

At the same time, many natural systems that display apparent rhythmicity are high-dimensional, strongly stochastic, and spectrally stable.
In such systems, eigenvalues alone provide an incomplete description of the dynamics.
Non-normal operators, whose eigenvectors are non-orthogonal, are now known to support large transient amplification of perturbations,
even when all eigenvalues lie deep in the stable half-plane \cite{trefethen2005spectra,farrell1996generalized,schmid2007nonmodal}.
This phenomenon has long been recognized in fluid mechanics, atmospheric dynamics, and control theory
\cite{trefethen2005spectra,farrell1996generalized,schmid2007nonmodal,trefethen1993hydrodynamic},
but its implications for collective temporal organization and synchronization-like behavior remain underexplored.

Recent work has shown that non-normality can generate pseudo-critical behavior \cite{troude2025illusion}:
sharp regime changes in macroscopic observables without any underlying bifurcation.
In these regimes, transient growth, dimensional reduction, and pronounced sensitivity to noise emerge despite spectral stability.
However, how such pseudo-critical amplification manifests in time-domain observables, how it shapes the spectral content of stochastic fluctuations, and whether it can generate apparent synchronization in the absence of oscillators remain open questions.

Here, we address these questions by introducing the concept of \emph{pseudo-coherence}. Pseudo-coherence denotes an emergent intermittent form of collective organization driven by non-normal stochastic amplification. It is characterized by transient phase alignment, dominance of reactive modes, and sustained entropy production, yet it lacks the hallmarks of classical synchronization: there are no pre-existing oscillators to synchronize, no intrinsic frequency, no stable spectral peak, and no underlying oscillatory instability. Instead, coherence arises geometrically from the structure of the pseudospectrum and the consequent redistribution of noise-induced fluctuations \cite{trefethen2005spectra}.

We first develop a minimal stochastic framework that isolates the role of non-normality.
Within this setting, we show analytically that increasing non-normality induces a pseudo-critical transition,
marked by the emergence of irreversible probability currents and non-zero Kuramoto-like order parameters,
while the system remains linearly stable.
We demonstrate that pseudo-critical non-normal dynamics reshapes the imaginary pseudospectrum,
amplifying low-frequency fluctuations and suppressing intermediate frequencies,
thereby producing apparent characteristic frequencies only under finite-time or localized spectral estimation.

The framework developed here has broad implications.
Non-normal operators are ubiquitous in ecological networks, climate systems, seismology, and other natural systems
where apparent oscillations and regime shifts are routinely observed \cite{allesina2012stability,sornette2004critical,trefethen2005spectra}.
Our results suggest that a significant fraction of such phenomena may reflect pseudo-coherent stochastic dynamics
rather than true oscillatory mechanisms or critical transitions.
By linking non-normal geometry, entropy production, and collective observables,
this work provides a new lens through which temporal organization in complex systems can be understood.

\section{Results}

\paragraph{\textbf{Model and Reduced Non-Normal Dynamics}}
We consider a linear overdamped stochastic system of dimension $N$,
\begin{equation}
\dot{\x}(t) = \A \x(t) + \xii(t),
\label{eq:VAR}
\end{equation}
where $\x(t) \in \mathbb{R}^N$ denotes the system state, $\A \in \mathbb{R}^{N \times N}$ is a constant interaction matrix,
and $\xii(t)$ is a vector of stochastic forcing terms. We will consider the general case where 
$\A$ is non-normal, i.e., it does not commutes with its transpose, $\A \A^\top \neq \A^\top \A$,
in which case its eigenvectors are non-orthogonal. This leads to interactions between 
the decays of different modes that may interfere constructively,
forming strong transient amplification even though the system is asymptotically stable.
This phenomenon is geometric rather than spectral and is invisible to eigenvalue-based stability analysis \cite{trefethen1993hydrodynamic,trefethen2005spectra}.
Non-normal amplification has been shown to induce pseudo-critical behavior in a wide range of systems,
generating large responses without proximity to genuine instabilities \cite{troude2025illusion,troude2025Unifying}.

Throughout this work, we impose three structural assumptions:
\begin{enumerate}
\item \emph{Linear stability}: all eigenvalues of $\A$ have strictly negative real parts.
\item \emph{Absence of intrinsic oscillators}: $\A$ has no complex-conjugate eigenvalues.
\item \emph{No periodic forcing}: $\xii(t)$ is temporally uncorrelated noise with zero mean.
\end{enumerate}
Thus, model \eqref{eq:VAR} contains \emph{no microscopic oscillatory units}, no preferred frequencies,
and no synchronizing interactions of the Kuramoto type \cite{kuramoto1975self,acebron2005kuramoto,pikovsky2003synchronization}.
Any emergence of coherent temporal organization can therefore only originate from the geometry of $\A$ itself, as we show below.

Under these conditions, system \eqref{eq:VAR} admits a unique stationary distribution.
When $\A$ is normal, the dynamics reduces to a multivariate Ornstein-Uhlenbeck process exhibiting purely relaxational dynamics.
Independent modes have their fluctuations decay monotonically.
Isotropic noise produces featureless, monotone power spectra.

A key consequence of strong non-normality is an effective reduction of the system's stochastic dynamics onto a low-dimensional subspace spanned by highly aligned eigenvectors.
This geometric dimensional reduction has been identified as a universal feature of non-normal systems operating in pseudo-critical regimes \cite{troude2025illusion}.
The two reduced directions play asymmetric dynamical roles.
The second component acts as a \emph{non-normal mode},
weakly excited by noise, while the first component defines a \emph{reaction mode} into which perturbations are transiently redirected and amplified.
This reaction-mode structure is the fundamental building block of non-normal amplification \cite{troude2025Unifying,troude2025illusion}
and can be expressed as follows.
Let $\PP = (\pp_1, \pp_2)$ be an isometry embedding a two-dimensional non-normal subspace into $\mathbb{R}^N$,
with $\pp_1$ and $\pp_2$ orthonormal.
Defining the reduced coordinates $\z = \PP^\top \x$, the projected dynamics reads
\begin{equation}
\dot{\z}(t) = \Gam \z(t) + \et(t),
\label{eq:reduced}
\end{equation}
where $\et(t) = \PP^\top \xii(t)$.
Ref.~\cite{troude2025Unifying}  showed that, up to a unitary transformation
(embedded in the isometry),
the matrix $\Gam$ can be written
\begin{equation}
\Gam =
\begin{pmatrix}
-\alpha & \beta \kappa \\
\beta / \kappa & -\alpha
\end{pmatrix},
\qquad \alpha > \beta > 0,\quad \kappa \geq 1 ~.
\label{eq:Gamma}
\end{equation}
Crucially, the eigenvalues of $\Gam$ remain real and strictly negative for all $\kappa$,
ensuring the complete absence of intrinsic oscillatory modes.
While $\alpha$ and $\beta$ control the overall relaxation rates and degenerency of the sub-system,
the dimensionless parameter $\kappa$ encodes the degree of non-normality,
i.e., the geometric alignment between eigenvectors:
$\kappa = 1$ corresponds to a normal system,
while $\kappa \gg 1$  implies pronounced eigenvector non-orthogonality, characterized by the quasi-alignment of eigenvector pairs.
This representation allow us to capture explicitly the non-normality of the system via the degree of non-normality $\kappa$.

Following \cite{troude2025Unifying}, we introduce the non-normality index
\begin{equation}
K = \frac{\kappa - \kappa^{-1}}{2},
\label{eq:K}
\end{equation}
which vanishes in the normal case and grows monotonically with eigenvector non-orthogonality.
This index quantifies the maximal transient amplification allowed by the linear dynamics.

A central result established in \cite{troude2025Unifying} is the existence of a \emph{pseudo-critical threshold} 
\begin{equation}
    K_c = \sqrt{\frac{\sqrt{1-\delta^2}}{1-\sqrt{1-\delta^2}}},
    \quad \delta = \left|\frac{\beta}{\alpha}\right| ;
\end{equation}
defined as the minimal non-normality such that, only for $K>K_C$ does
transient amplification reshape stochastic relaxation.
Importantly, this threshold does \emph{not} correspond to a spectral instability:
the system remains linearly stable for all $K$, and no eigenvalue approaches the imaginary axis.
We emphasize that pseudo-criticality refers here to a geometric transition in transient dynamics,
not to a bifurcation or phase transition, and
there is no mechanism for phase locking in the classical sense \cite{kuramoto1975self,acebron2005kuramoto,pikovsky2003synchronization,glass1988clocks}.
\newline

\paragraph{\textbf{Emergence of Pseudo-Coherence and pseudo-synchronization without oscillators}}
The projection of the dynamics onto the non-normal subspace (\ref{eq:reduced}) induces a geometrically structured organization in two clusters.
In the pseudo-critical regime, the dynamics of $\x(t)$ is dominated by the reaction mode $\x(t) \approx z_1(t)\,\pp_1$
where $\pp_1$ is the reaction-mode vector introduced to obtain (\ref{eq:reduced}).
The sign structure of $\pp_1$ naturally partitions the system into clusters, reflecting coordinated yet oppositely directed contributions among components:
\begin{equation}
\mathcal{C}_+ = \{ i \mid (\pp_1)_i > 0 \}, 
\qquad
\mathcal{C}_- = \{ i \mid (\pp_1)_i < 0 \},
\label{eq:clusters}
\end{equation}
with a possible residual set where $(\pp_1)_i \approx 0$.
These clusters are \emph{purely geometric}: they are determined by the structure of the interaction matrix, not by any dynamical symmetry, coupling topology, or oscillator subpopulation. 

To quantify the emergence of collective alignment, we associate to each component $x_i(t), i=1,..,N$, an instantaneous phase $\theta_i(t)$ extracted from the stochastic time series using a non-parametric procedure (see Methods).
No assumption of sinusoidal oscillations or intrinsic periodicity is made.
A Kuramoto-like order parameter is defined by
\begin{equation}
R(t) = \left| \frac{1}{|\mathcal{C}|} \sum_{i \in \mathcal{C}} e^{\mathrm{i}\theta_i(t)} \right|,
\label{eq:order_parameter}
\end{equation}
computed either over the full system or restricted to a given cluster $\mathcal{C}_\pm$.
We stress that \eqref{eq:order_parameter} is used here \emph{exclusively as a measurement tool}.
The underlying dynamics does not contain oscillators, phase variables,
coupling functions, or frequency distributions of the Kuramoto type.

\begin{figure*}
    \centering
    \includegraphics[width=\textwidth]{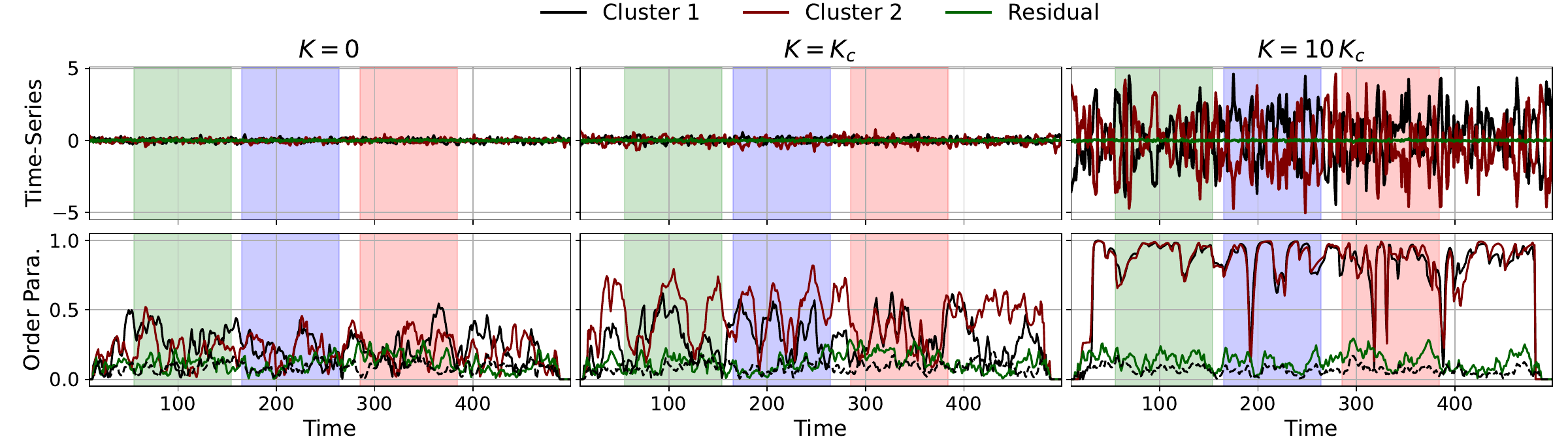}
    \caption{
        \textbf{Emergent coherence and synchronization induced by non-normality in an overdamped stochastic system without oscillators.} 
        Synthetic time series \eqref{eq:VAR} for $N=100$,  $\alpha=1$, $\beta=0.1$, and $\epsilon=10^{-3}$ (top row) and corresponding Kuramoto-like order parameters \eqref{eq:order_parameter} (bottom row) for three values of the non-normality index (\ref{eq:K}):
        normal dynamics ($K=0$, left), pseudo-critical regime ($K=K_c$, center), and strongly non-normal regime ($K=10K_c$, right).
        The system is purely overdamped, linearly stable, driven by uncorrelated noise, and contains no intrinsic oscillators or periodic forcing.
        The time series are projected onto three geometric groups defined by the sign structure of the reaction mode \eqref{eq:clusters}:
        two non-normal clusters (Cluster~1 and Cluster~2) and a residual subspace orthogonal to the non-normal sub-space. The dashed black line
        corresponds to the global order parameter.
        Colored background bands indicate successive time windows used as reference in Figure \ref{fig:synthetic_spec}.
        }
    \label{fig:synthetic_series}
\end{figure*}

Figure~\ref{fig:synthetic_series} shows how increasing the non-normality index $K$ transforms the system's behavior.
For $K<K_c$, the dynamics are effectively normal: the system reduces to a multivariate Ornstein--Uhlenbeck process, phases are uncorrelated, and the global order parameter $R(t)$ fluctuates near zero without collective organization.
As $K$ approaches and exceeds $K_c$, intermittent bursts of large order parameter emerge within the geometric clusters $\mathcal{C}_\pm$. These episodes reflect strong but transient phase alignment. In contrast, the order parameter computed over the full system remains small, due to the coexistence of synchronized and anti-synchronized components imposed by the sign structure of the reaction mode, while orthogonal components remain largely uninvolved. In the strongly non-normal regime, coherence becomes more pronounced but remains intermittent and non-stationary.

When the reaction mode transiently amplifies, components in $C_+$ grow coherently, whereas components in $C_-$ decrease coherently. This opposition can resemble anti-synchronization, yet both clusters simply co-evolve along the same reaction direction. Any increase in $R(t)$ therefore reflects a collective excursion along this geometric mode rather than coupling between microscopic oscillators.
More generally, the observed synchronization-like phenomenology emerges in a purely overdamped, linearly stable system driven by uncorrelated noise, with no intrinsic oscillators, characteristic frequencies, or phase coupling. It disappears in the normal limit $K \to 0$, demonstrating that non-normality alone is sufficient to produce intermittent, noise-amplified geometric alignment.

\begin{figure}
    \centering
   \includegraphics[width=0.48\textwidth, trim=0 0 20.5cm 0,clip]{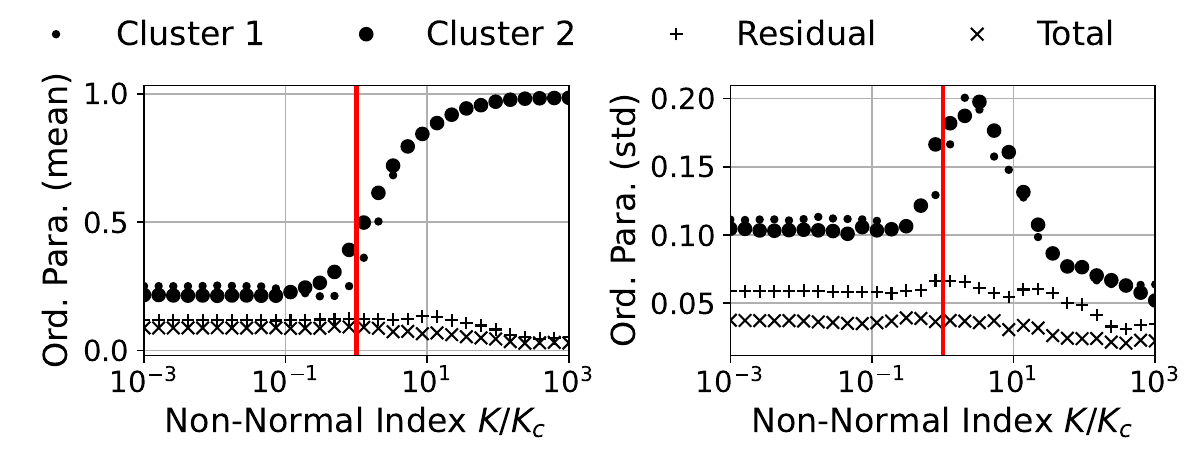}
    \caption{
        \textbf{Non-normality-driven transition.} 
        Mean (left panel) and standard deviation (right panel) of the Kuramoto-like order parameters \eqref{eq:order_parameter} of the reaction clusters, the residual subspace, and the full system  as a function of the normalized non-normality index $K/K_c$ ($N=100$, $\alpha=1$, $\beta=0.1$, $\epsilon=10^{-3}$). 
The residual and global measures remain nearly constant, while the reaction clusters exhibit a sharp increase in mean and a pronounced peak in variance near $K/K_c \sim 1$, marking the onset of intermittent collective organization.     }
    \label{fig:phase_transition}
\end{figure}

Building on Figure~\ref{fig:synthetic_series}, which illustrates representative time series and cluster order parameters for selected values of $K$, Figure~\ref{fig:phase_transition} quantifies this behavior systematically by plotting 
the mean and standard deviation of the order parameters of different clusters as functions of the control parameter $K/K_c$. 
It reveals the existence of a well-defined non-normality-driven transition.
For $K \ll K_c$, the system behaves as an effectively normal multivariate Ornstein--Uhlenbeck process
where  cluster order parameters remain small.
As $K/K_c$ approaches unity, a collective reorganization sets in. The order parameter within the reaction clusters increases markedly and becomes strongly intermittent
as evidenced by the large peak in the standard deviation.
Throughout this transition, all eigenvalues remain strictly stable: no Hopf bifurcation or spectral instability occurs. 
Instead, sufficiently strong transient amplification channels stochastic fluctuations into the reaction mode, which acquires extensive system-wide support. The resulting regime is collective, temporally structured, and irreversible despite purely linear, overdamped dynamics. We interpret this coordinated crossover as a \emph{non-normal phase transition}, governed by pseudospectral amplification \cite{trefethen1993hydrodynamic,trefethen2005spectra} rather than eigenvalue criticality.

\begin{figure*}
    \centering
    \includegraphics[width=\textwidth]{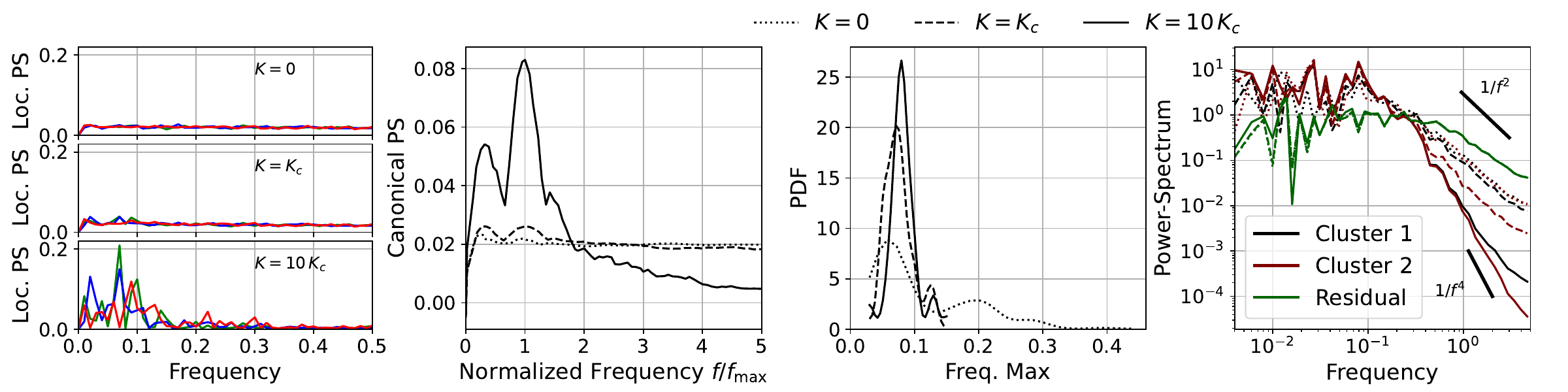}
    \caption{
        \textbf{ Spectral analysis of the synthetic time series of Fig.~\ref{fig:synthetic_series} for three values of the non-normality index ($K=0$, $K=K_c$, $K=10K_c$).} \emph{Left panel:} Power spectra computed over three typical time windows of fixed length (100 data points); each curve corresponds to a different window.
\emph{Center-left:} Canonical averaged spectrum $\hat P(f/f_{\max})$  obtained by rescaling frequencies by the peak frequency of each window and averaging across windows (Savitzky-Golay smoothing applied for visualization).
\emph{Center-right:} Distribution of the peak frequencies $f_{\max}$  over a large ensemble of time windows.
\emph{Right panel:} Power spectra computed over the full time series for the reaction clusters and the residual subspace, using FFT and averaging over components within each cluster. For $K=0$ and $K=K_c$, the tail of the spectrum is proportional to $1/f^2$, compared to to $1/f^4$ for $K=10 K_c$.
    }
    \label{fig:synthetic_spec}
\end{figure*}

We now connect the emergence of pseudo-coherence and cluster synchronization
(Figs.~\ref{fig:synthetic_series} and \ref{fig:phase_transition})
to their spectral signatures.
Figure~\ref{fig:synthetic_spec} analyzes the power spectra computed from the same synthetic time series of Fig.~\ref{fig:synthetic_series}
for increasing non-normality ($K/K_c=0,1,10$),
separately for the reaction clusters and the residual subspace.

In the normal regime ($K=0$), spectra are consistent with a multivariate Ornstein--Uhlenbeck process:
broadband, monotonic, and governed by a single relaxation scale.
No frequency stands out, reflecting purely overdamped, reversible dynamics.
As $K$ approaches and exceeds $K_c$, spectral weight reorganizes strongly along the reaction clusters.
Power concentrates at low frequencies and the spectrum develops an extended algebraic regime
close to a $1/f^4$ decay in the strongly non-normal case (Fig.~\ref{fig:synthetic_spec}, right panel), as predicted theoretically (see Supplementary Material).
This reshaping reflects transient amplification along the reaction mode,
which integrates stochastic fluctuations and generates long temporal correlations.
The residual subspace, by contrast, retains its OU-like spectral structure.

Finite-time spectral analysis reveals how this collective organization manifests dynamically.
When spectra are computed over successive time windows (Fig.~\ref{fig:synthetic_spec}, left panel),
well-defined interior peaks systematically emerge in the reaction clusters once $K/K_c \gtrsim 1$.
These peaks correspond to finite global frequencies organizing the fluctuations during each window.
Their peak frequency $f_{\max}$ drifts gradually in time,
yet within each window the spectral concentration remains sharp and highly coherent.
The distribution of the peak frequencies $f_{\max}$ (Fig.~\ref{fig:synthetic_spec}, center-right)
over a large ensemble of time windows narrows markedly as non-normality increases,
indicating the progressive concentration of spectral power around finite frequencies.

To highlight this emergence of global coherent oscillatory modes, we rescale each windowed spectrum by its peak frequency $f_{\max}$
and construct canonical averaged normalized spectra $\hat P(f/f_{\max})$ \cite{can-ave97,JohSor-can98}
(Fig.~\ref{fig:synthetic_spec}, center-left). Canonical averaging is performed using quantile binning across windows,
followed by a Savitzky--Golay filter applied to $\hat{P}$ to enhance readability.
This averaged spectrum reveals a coherent structure centered around the dominant frequency of each episode.
These frequencies are not fixed dynamical eigenmodes but reflect the effective timescale of the transient amplification bursts that structure the collective dynamics.
In the Supplementary Material, the Fourier analysis of Fig.~\ref{fig:synthetic_spec} is complemented and strengthened with a time-frequency analysis based on the Morlet wavelet transform.

The resulting picture is therefore not one of static oscillators,
but of dynamically generated global frequencies.
Non-normal amplification funnels fluctuations into the reaction mode,
producing coherent stochastic excursions with finite effective durations.
Each episode carries a dominant frequency set by the duration and intensity of the amplification burst,
and the sequence of such events generates slowly drifting but highly coherent spectral peaks.
Non-normality thus produces a genuine form of spectral organization:
not through eigenvalue instability or intrinsic oscillators,
but through geometry-induced transient amplification
that spontaneously creates coherent system-wide frequency structure in a purely linear,
overdamped, noise-driven system.

We term this phenomenon \emph{pseudo-coherence}. Extending the notions of pseudospectrum \cite{trefethen2005spectra} and pseudo-criticality \cite{troude2025illusion}, pseudo-coherence describes the emergence of temporally organized and spectrally structured collective behavior in systems that contain neither intrinsic oscillators nor critical eigenvalues. It originates from the imaginary pseudospectrum of the Jacobian, which controls how stochastic fluctuations are redistributed across time scales. Non-normality selectively amplifies slow modes while preserving overall stability. Once amplification becomes sufficiently strong, the reaction mode acquires extensive support across system components, concentrating fluctuations onto a low-dimensional subspace where they organize into coherent clusters, generate localized spectral bands, and produce persistent lead-lag asymmetries.

Within this framework, synchronization is not a primary dynamical mechanism but a secondary manifestation of pseudo-coherence. As non-normality increases beyond the transition, the reaction mode induces intermittent phase alignment across components. Elements with opposite signs in the reaction mode naturally appear anti-synchronized even though they are driven by the same stochastic collective mode. This geometric sign structure explains the coexistence of synchronized and anti-synchronized clusters without invoking competing oscillators \cite{glass1988clocks}, chimera states \cite{abrams2004chimera}, or explosive synchronization \cite{gomez2011explosive}. Kuramoto-like order parameters therefore measure instantaneous projection onto the reaction subspace rather than phase locking between intrinsic oscillators.
\vskip 0.2cm

    \paragraph{\textbf{Time Reversal Symmetry Breaking and Entropy Production.}}
    Pseudo-coherence is not limited to transient phase alignment and finite-time spectral organization.
    Its onset is also marked by the emergence of an intrinsic arrow of time and by the appearance of a non-vanishing thermodynamic cost,
  which are direct signatures of the same non-normal amplification mechanism that generates pseudo-synchronization. 
    To make this explicit, we first introduce an empirical measure of temporal asymmetry,
    then show that it undergoes a sharp transition at the onset of pseudo-coherence,
    and finally connect this transition to the entropy production rate of the reduced non-normal dynamics. 
    
    \begin{figure}
    \centering
    \includegraphics[width=0.3\textwidth, trim=20cm 0 11cm 0, clip]{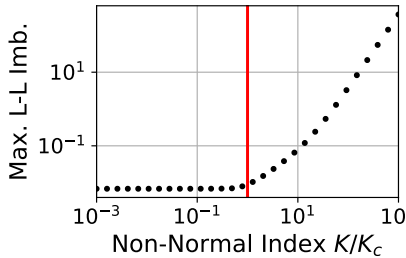}
    \caption{Maximum (over $\tau\in[0,10]$) of the lead-lag imbalance measure \eqref{eq:main_ll_def} as a function of $K/K_c$.}
    \label{fig:phase_entropy}
\end{figure}

    A natural way to test whether a stochastic dynamics is reversible is to compare how fluctuations propagate forward versus backward in time. 
    For the reduced process \eqref{eq:reduced}, we define the stationary lagged covariance matrix
  $\C(\tau)=\lim_{t\to\infty}\mw{\x(t)}{\x(t+\tau)^\top}$,
    and its antisymmetric component
   $ \Delta\C(\tau)=\frac{1}{2}\big(\C(\tau)-\C(\tau)^\top\big)$,
    using the identity $\C(-\tau)=\C(\tau)^\top$. 
    If the dynamics is time-reversal symmetric, then forward and backward lagged correlations coincide and $\Delta\C(\tau)=0$. 
    This motivates the scalar lead--lag imbalance measure
    \begin{equation}
    I(\tau)=\frac{1}{\sqrt{2}}\|\C(\tau)-\C(\tau)^\top\|_F,
    \label{eq:main_ll_def}
    \end{equation}
    which provides a direct empirical quantification of the arrow of time. 
    It vanishes if and only if time-reversal symmetry holds at lag $\tau$,
    while positive values indicate that fluctuations lead and lag asymmetrically and therefore reveal irreversible dynamics. 
    
    Figure~\ref{fig:phase_entropy} shows that the lead-lag imbalance measure exhibits a clear
    transition at $K=K_c$ from values close to zero for $K\ll K_c$ (reversible fluctuations)
    to increasing imbalance for $K > K_c$, marking the onset of a macroscopic arrow of time. 
    Thus, the same threshold that organizes fluctuations into pseudo-coherent collective episodes also marks the appearance of
    an irreversible temporal behavior.
    
    This result can be derived precisely within the reduced non-normal dynamics. 
    We consider the stochastic forcing in \eqref{eq:reduced} such that
    \(
    \mw{\et(t)}{\et(t')^\top}=\B\,\delta(t-t'),
    \)
    with covariance matrix
    \begin{equation}
    \B=
    \begin{pmatrix}
    \sigma_1^2 & \rho\,\sigma_1\sigma_2 \\
    \rho\,\sigma_1\sigma_2 & \sigma_2^2
    \end{pmatrix}.
    \label{eq:main_B_def}
    \end{equation}
    Here $\sigma_i^2=\mw{\eta_i(t)^2}$ denotes the variance of the noise along reduced direction $i$,
    and $\rho$ is the correlation coefficient defined by
    \(
    \mw{\eta_1(t)\eta_2(t')}=\rho\,\sigma_1\sigma_2\,\delta(t-t').
    \)
    For this reduced dynamics, the lead--lag imbalance (\ref{eq:main_ll_def}) takes the explicit form
    \begin{equation}
    I(\tau)=
    \frac{\sigma_1\sigma_2}{2\sqrt{2}\alpha}
    |K_\sigma|
    \left|e^{-(\alpha-\beta)\tau}-e^{-(\alpha+\beta)\tau}\right|,
    \label{eq:main_ll_closed}
    \end{equation}
    where
   $
    K_\sigma=\frac{1}{2}\Big(\kappa_\sigma-\kappa_\sigma^{-1}\Big),
    \kappa_\sigma=\kappa\,\frac{\sigma_2}{\sigma_1}.
 $
    Equation~\eqref{eq:main_ll_closed} shows that the arrow of time is directly controlled by the same effective non-normality parameter $K_\sigma$ that governs amplification in the reduced dynamics. 
    If $K_\sigma=0$, then $I(\tau)=0$ for all $\tau$ and the process is reversible. 
    If $K_\sigma\neq0$, then $I(\tau)$ becomes positive over a finite range of lags, revealing a preferred temporal direction. 
    In the isotropic-noise case, where $\sigma_1=\sigma_2$ and $\rho=0$, one has $\kappa_\sigma=\kappa$ and therefore $K_\sigma=K$. 
    The lead--lag imbalance is then controlled purely by the geometric non-normality of the system. 
    The sharp increase seen in Fig.~\ref{fig:phase_entropy} therefore reflects the onset of irreversible dynamics induced by non-normal amplification, rather than any spectral instability. \\
    This temporal asymmetry is the observable signature of a deeper non-equilibrium structure. 
    A non-zero antisymmetric component of the lagged covariance implies broken detailed balance and is associated with circulating probability currents in the reduced subspace. 
    In a stationary non-equilibrium steady state (NESS), the thermodynamic cost of maintaining these currents is quantified by the entropy production rate $\Phi$. 
    For linear Gaussian systems of the form \eqref{eq:reduced}, $\Phi$ can be written in terms of the left and right eigenvectors $\lv_i$, $\rr_i$ associated with the eigenvalues $-\lambda_i$ of $\Gam$ as
    \begin{equation}
    \Phi=\sum_{i,j}
    \left(\lv_i\!\cdot\!\B\lv_j\right)
    \left(\rr_j\!\cdot\!\B^{-1}\rr_i\right)
    \left[-\lambda_i\frac{\lambda_i-\lambda_j}{\lambda_i+\lambda_j}\right].
    \end{equation}
    In our reduced non-normal setting, the eigenvalues are real and negative, so only the cross-terms contribute. 
    After explicit evaluation (see Supplementary Material), the entropy production rate takes the closed form
    \begin{equation}
    \Phi=\frac{2\beta^2}{\alpha}\,
    \frac{K_\sigma^2}{1-\rho^2}.
    \label{eq:main_EPR_final}
    \end{equation}
    Equation~\eqref{eq:main_EPR_final} makes explicit that entropy production is governed by the same effective non-normality parameter $K_\sigma$ that controls the lead--lag imbalance. 
    It vanishes in the normal limit and grows quadratically as non-normality increases. 
    Thus, in the pseudo-coherent regime, the lead--lag imbalance is not merely a phenomenological indicator of temporal asymmetry: it is a direct empirical manifestation of the entropy-producing probability currents that sustain the non-equilibrium steady state. 
    The emergence of an arrow of time and the onset of entropy production are therefore two complementary expressions of the same irreversible organization. \\
    Taken together, these results establish the thermodynamic meaning of pseudo-coherence. 
    Non-normal amplification concentrates stochastic fluctuations onto the reaction mode, where they become sufficiently strong to organize transient phase alignment, finite-time spectral structure, and asymmetric lead--lag correlations. 
    At the same time, this amplification breaks detailed balance, generates circulating probability currents, and produces a strictly positive entropy production rate. 
    The system therefore settles into a non-equilibrium steady state whose thermodynamic cost reflects the maintenance of the same collective dynamics that underlies pseudo-synchronization. \\
    Pseudo-coherence thus defines a resilient mechanism for collective synchronization in stable overdamped systems. 
    Because all eigenvalues remain strictly in the stable half-plane, the system preserves linear stability while sustaining irreversible collective dynamics. 
    Temporal directionality, entropy production, and apparent synchronization all emerge from the same geometric mechanism: non-normal pseudospectral amplification. 
    In this sense, pseudo-coherence extends synchronization theory beyond oscillator-based frameworks and identifies non-normal geometry as a generic route to temporal order in complex systems.


\section{Discussion}


A promising application of pseudo-coherence concerns large-scale brain dynamics. 
Neural activity is commonly organized into canonical frequency bands (delta (0.5-4 Hz), theta (4-8 Hz), alpha (8-12 Hz), beta (13-30 Hz), and gamma (30-100 Hz)) often interpreted as signatures of oscillatory circuits or near-critical synchronization \cite{Buzsaki2004,Buzsaki2006,Buzsaki2012}. Within the pseudo-coherent framework, such rhythms can arise naturally as finite-time spectral concentrations generated by non-normal amplification in spectrally stable networks. Because these rhythms are typically transient and context-dependent, their dynamics are consistent with the pseudo-coherence mechanism. Transient growth along dominant reaction modes organizes stochastic fluctuations into coherent episodes with well-defined effective time scales, producing drifting yet sharply defined spectral peaks within these characteristic bands without requiring intrinsic oscillators or finely tuned criticality.
This perspective is closely related to Buzs\'aki's ``rhythm of the brain'' framework, which emphasizes that large-scale rhythms arise from network-level coordination across multiple spatial and temporal scales \cite{Buzsaki2006,Buzsaki2012}. 
Pseudo-coherence provides a complementary dynamical mechanism for this organization: strongly non-normal neural circuits can transform broadband fluctuations into coherent frequency bands through pseudospectral amplification, potentially explaining the coexistence of robustness, variability, and intermittent synchronization observed in neural recordings.

Another promising domain of application of the pseudo-coherence framework may be the dynamics of genome-resolved gut microbiomes. 
Recent advances in longitudinal shotgun metagenomics now allow microbial communities to be tracked at genome resolution through metagenome-assembled genomes (MAGs), enabling the temporal dynamics of individual microbial populations to be reconstructed directly from time-series metagenomic data \cite{orellana2023timeseriesMAG,cheng2024genomeresolved}, revealing 
intermittent oscillation-like patterns and clusters of apparently synchronized taxa \cite{microbiome2025}. Because microbial genomes are not intrinsic oscillators and the system is highly stochastic and high-dimensional, such observations are not easily interpreted within classical synchronization frameworks. We hypothesize that these patterns could reflect pseudo-coherent dynamics: transient amplification of stochastic fluctuations along non-normal reaction modes in an otherwise stable community network. In this view, intermittent synchronization and drifting spectral features would arise from collective amplification distributed across many taxa rather than from intrinsic oscillators or strict circadian entrainment. Testing this hypothesis in microbiome time series may help determine whether non-normal pseudo-coherence provides a useful dynamical explanation for the coordinated yet highly variable behavior observed in microbial ecosystems.

In this work, we identified a new route to collective temporal organization in stochastic dynamical systems that does not rely on intrinsic oscillators, phase coupling, or proximity to spectral instabilities. We showed that \emph{non-normal pseudospectral amplification} in linearly stable systems can generate \emph{pseudo-coherence}: a regime of intermittent, synchronization-like behavior emerging as a sharp transition when the non-normal control parameter crosses a pseudo-critical threshold. In this regime, stochastic fluctuations concentrate along a dominant reaction mode, producing coherent clusters, asymmetric lead--lag correlations, circulating probability currents, and strictly positive entropy production, despite the absence of eigenvalue crossings or Hopf bifurcations. A central result is that pseudo-critical non-normal dynamics reshapes the imaginary pseudospectrum, strongly amplifying slow fluctuations and generating finite-time spectral organization that appears as drifting yet coherent frequency bands without intrinsic oscillators. These findings establish pseudo-coherence as a new category of collective behavior, namely geometric, stochastic, and inherently non-equilibrium, providing a unified explanation for how organized temporal patterns and apparent rhythms can arise in stable, high-dimensional systems. Because non-normal interactions are ubiquitous in complex networks, this mechanism may underlie a wide range of observed oscillatory or synchronization-like phenomena across fields ranging from neuroscience and ecology to climate dynamics.


\section{Materials and Methods}


All the derivations are in the Supplementary Material.
\newline

\paragraph{\textbf{Numerical set-up}}
To validate our theoretical predictions,
we generate synthetic data from an $N$-dimensional Vector Auto-Regressive (VAR) process of the form
\begin{equation}
    \begin{split}
        &\x_{t+\delta t} = \A(\delta t)\x_t + \sqrt{\delta t}\,\xii_t, \\
        &\A(\delta t) = \PP_N^\dag \left(\I + \delta t\,\Gam\right)\PP_N + \epsilon\,\delta t\,\E, \\
        &\E = \frac{1}{\sqrt{2N}}\left(\E' + \E'^\dag\right),
        \qquad E'_{ij}\sim\mathcal{N}(0,1).
    \end{split}
\end{equation}
Here $\xii_t$ denotes a vector of independent stochastic increments,
drawn from identical Poisson processes and rescaled by $\sqrt{\delta t}$ to ensure a well-defined continuous-time limit.

The matrix $\Gam$ specifies the dynamics within the reduced two-dimensional non-normal subspace and is given by
\begin{equation}
    \Gam =
    \begin{pmatrix}
        -\alpha & \beta\kappa \\
        \beta/\kappa & -\alpha
    \end{pmatrix}.
\end{equation}
The matrix $\PP_N=(\pp_1\,,\,\pp_2)$ is an isometry embedding this reduced subspace into the full $N$-dimensional space. The direction $\pp_1$ corresponds to the reaction mode, while $\pp_2$ defines the associated non-normal mode.

The parameter $\kappa\ge 1$ controls the degree of non-normality: the system is normal for $\kappa=1$ and becomes increasingly non-normal as $\kappa$ grows. We adopt the convention $\beta\ge 0$, and linear stability of the reduced dynamics is ensured whenever $\alpha>\beta$.

Following \cite{troude2025Unifying}, we introduce the non-normality index
\begin{equation}
    K = \frac{\kappa-\kappa^{-1}}{2},
\end{equation}
which quantifies the geometric amplification associated with non-orthogonal modes. A system is said to be \emph{pseudo-critical} when it is linearly stable and the non-normality exceeds a critical threshold,
\begin{equation}
    K \ge K_c,
    \qquad
    K_c = \sqrt{\frac{\sqrt{1-\delta^2}}{1-\sqrt{1-\delta^2}}},
    \qquad
    \delta = \frac{\beta}{\alpha}.
\end{equation}
This threshold is crucial, as it marks the regime in which each noise increment ultimately decays exponentially, yet undergoes transient amplification before doing so. In other words, the kernel through which the noise propagates is not monotonically decaying, giving rise to transient, or local, instability \cite{troude2025illusion}.
All numerical simulations are performed with a time step $\delta t = 0.1$ over a total duration $T=500$, corresponding to $5000$ sampled data points. Unless otherwise stated, we fix the parameters to $\alpha=1$, $\beta=0.1$, and $\epsilon=10^{-3}$.

The non-normal mode and reaction used takes the form 
\begin{equation}
    \begin{split}
        &\pp_{1,i} = 
        \begin{cases}
            \frac{1}{\sqrt{50}} , & i=1,\ldots,25, \\
            -\frac{1}{\sqrt{50}} , & i=26,\ldots,50, \\
            0 , & i=51,\ldots,100,
        \end{cases}
        \\
        &\pp_{2,i} =
        \begin{cases}
            \frac{1}{\sqrt{50}} , & i=1,\ldots,50, \\
            0 , & i=51,\ldots,100,
        \end{cases}
    \end{split}
\end{equation}
so that they are orthogonal and share identical support, ensuring that their support metrics are equal to $s(\pp_1)=s(\pp_2)=\frac{1}{\sqrt{2}}$.
The two anti-synchronized clusters are therefore of equal size, each spanning $25$ dimensions, while the remaining $50$
dimensions do not participate in the non-normal dynamics. This setup illustrates a situation in which only a subset of the system is driven by the reaction mode, whereas the rest evolves independently of it. This construction is motivated by metagenome-assembled genome (MAG) data from bacterial communities in the mouse gut, which suggest that collective behavior may emerge within a subset of components while others remain uninvolved \cite{microbiome2025}.
\newline

\paragraph{\textbf{Phase extraction from time-series data}}
To extract an instantaneous phase from noisy and non-sinusoidal time-series data,
we adopt a non-parametric procedure based on local trend detrending and statistical detection of directional transitions,
following the same procedure as the one proposed in \cite{microbiome2025}.
This approach does not assume a fixed waveform or global periodicity and is therefore well suited for irregular and noisy empirical signals.

\paragraph{Median detrending.}
Given a discrete time series $x(t)$ sampled at uniform intervals,
slow baseline drifts were first removed using a moving median filter.
Specifically, we define the local median trend as
\begin{equation}
    x_{\mathrm{med}}(t) =
    \mathrm{Median}\{x(t-12),x(t-11), \ldots, x(t+11)\},
\end{equation}
corresponding to a 24-hour window at hourly resolution.
Boundary points and missing values were ignored when computing the median.

The detrended signal was then defined as
\begin{equation}
    x'(t) = x(t) - x_{\mathrm{med}}(t),
\end{equation}
which removes slow variations while preserving short-term oscillatory fluctuations.

\paragraph{Local directional significance via Fisher's exact test.}
To quantify whether the signal at time $t$ is locally increasing or decreasing,
we compared the signs of detrended fluctuations in the preceding and following time windows.
First, a local reference level was defined as
\begin{equation}
    m(t) =
    \mathrm{Median}\{x'(t-12), x'(t-11), \ldots, x'(t+11)\}.
\end{equation}

We then counted the number of points above and below this reference in the two windows:
\begin{align}
    a(t) &= \#\{x'(t-k) > m(t)\}, \quad k=1,\ldots,12, \\
    b(t) &= \#\{x'(t+k) > m(t)\}, \quad k=0,\ldots,11, \\
    c(t) &= \#\{x'(t-k) \le m(t)\}, \quad k=1,\ldots,12, \\
    d(t) &= \#\{x'(t+k) \le m(t)\}, \quad k=0,\ldots,11.
\end{align}

These counts define a $2\times2$ contingency table,
\begin{equation}
    \begin{pmatrix}
        a(t) & b(t) \\
        c(t) & d(t)
    \end{pmatrix},
\end{equation}
on which a one-tailed Fisher's exact test was performed.
The resulting $p$-value, denoted $p_{\mathrm{Fisher}}(t)$,
quantifies whether the distribution of signs differs significantly between the past and future windows.

A signed local significance score was then defined as
\begin{equation}
    w(t) = \mathrm{sign}\!\left[a(t)d(t)-b(t)c(t)\right]\,
    \log p_{\mathrm{Fisher}}(t).
\end{equation}
Positive values of $w(t)$ indicate statistically significant upward transitions,
whereas negative values indicate downward transitions.
Alternating positive and negative extrema of $w(t)$ separated by approximately half a period indicate robust oscillatory behavior.

\paragraph{Phase reconstruction from the local significance score.}
The temporal phase was reconstructed from the zero-crossings and local extrema of the $w(t)$ signal.
To reduce high-frequency noise, $w(t)$ was first smoothed using a short moving-average filter.
Zero-crossings were identified as changes in sign between consecutive time points,
with crossing times refined by linear interpolation.

Within each interval bounded by consecutive zero-crossings,
a local extremum of $w(t)$ was identified: a local maximum for positive intervals and a local minimum for negative intervals.
These extrema serve as anchor points for defining the oscillatory phase.

Each oscillatory cycle was divided into four segments:
\begin{enumerate}
    \item positive-to-negative zero-crossing ($\phi_T = 0$),
    \item local trough ($\phi_T = \pi/2$),
    \item negative-to-positive zero-crossing ($\phi_T = \pi$),
    \item local peak ($\phi_T = 3\pi/2$).
\end{enumerate}
Between consecutive anchor points, the phase was interpolated linearly in time.
Successive cycles were concatenated by adding $2\pi$ to ensure a monotonically increasing unwrapped phase $\phi_T(t)\in\mathbb{R}$.

The instantaneous phase was finally obtained by wrapping the unwrapped phase into the principal interval,
\begin{equation}
    \theta(t) = \phi_T(t) \bmod 2\pi ,
\end{equation}
with $\theta(t)\in[0,2\pi)$.

This procedure yields a continuous and robust phase estimate that is insensitive to waveform shape and long-term trends,
enabling reliable phase-based analyses of empirical time-series data.

\noindent
{\bf Acknowledgements}:  D.S. was partially supported by the National Natural Science Foundation of China (Grant No. T2350710802 and No. U2039202), Shenzhen Science and Technology Innovation Commission Project (Grants No. GJHZ20210705141805017 and No. K23405006), and the Center for Computational Science and Engineering at Southern University of Science and Technology.


\bibliographystyle{naturemag}   
\bibliography{bibliography}

\begin{thebibliography}{10}
\expandafter\ifx\csname url\endcsname\relax
  \def\url#1{\texttt{#1}}\fi
\expandafter\ifx\csname urlprefix\endcsname\relax\def\urlprefix{URL }\fi
\providecommand{\bibinfo}[2]{#2}
\providecommand{\eprint}[2][]{\url{#2}}

\bibitem{strogatz2003sync}
\bibinfo{author}{Strogatz, S.~H.}
\newblock \emph{\bibinfo{title}{Sync: The Emerging Science of Spontaneous
  Order}} (\bibinfo{publisher}{Hyperion}, \bibinfo{year}{2003}).

\bibitem{pikovsky2003synchronization}
\bibinfo{author}{Pikovsky, A.}, \bibinfo{author}{Rosenblum, M.} \&
  \bibinfo{author}{Kurths, J.}
\newblock \emph{\bibinfo{title}{Synchronization: A Universal Concept in
  Nonlinear Sciences}} (\bibinfo{publisher}{Cambridge University Press},
  \bibinfo{year}{2001}).

\bibitem{kuramoto1984chemical}
\bibinfo{author}{Kuramoto, Y.}
\newblock \emph{\bibinfo{title}{Chemical Oscillations, Waves, and Turbulence}}
  (\bibinfo{publisher}{Springer}, \bibinfo{year}{1984}).

\bibitem{stanley1999scaling}
\bibinfo{author}{Stanley, H.~E.}
\newblock \bibinfo{title}{Scaling, universality, and renormalization: Three
  pillars of modern critical phenomena}.
\newblock \emph{\bibinfo{journal}{Reviews of Modern Physics}}
  \textbf{\bibinfo{volume}{71}}, \bibinfo{pages}{S358--S366}
  (\bibinfo{year}{1999}).

\bibitem{trefethen2005spectra}
\bibinfo{author}{Trefethen, L.~N.} \& \bibinfo{author}{Embree, M.}
\newblock \emph{\bibinfo{title}{Spectra and Pseudospectra}}
  (\bibinfo{publisher}{Princeton University Press}, \bibinfo{year}{2005}).

\bibitem{farrell1996generalized}
\bibinfo{author}{Farrell, B.~F.} \& \bibinfo{author}{Ioannou, P.~J.}
\newblock \bibinfo{title}{Generalized stability theory. part i: Autonomous
  operators}.
\newblock \emph{\bibinfo{journal}{Journal of the Atmospheric Sciences}}
  \textbf{\bibinfo{volume}{53}}, \bibinfo{pages}{2025--2040}
  (\bibinfo{year}{1996}).

\bibitem{schmid2007nonmodal}
\bibinfo{author}{Schmid, P.~J.}
\newblock \bibinfo{title}{Nonmodal stability theory}.
\newblock \emph{\bibinfo{journal}{Annual Review of Fluid Mechanics}}
  \textbf{\bibinfo{volume}{39}}, \bibinfo{pages}{129--162}
  (\bibinfo{year}{2007}).

\bibitem{trefethen1993hydrodynamic}
\bibinfo{author}{Trefethen, L.~N.}, \bibinfo{author}{Trefethen, A.~E.},
  \bibinfo{author}{Reddy, S.~C.} \& \bibinfo{author}{Driscoll, T.~A.}
\newblock \bibinfo{title}{Hydrodynamic stability without eigenvalues}.
\newblock \emph{\bibinfo{journal}{Science}} \textbf{\bibinfo{volume}{261}},
  \bibinfo{pages}{578--584} (\bibinfo{year}{1993}).

\bibitem{troude2025illusion}
\bibinfo{author}{Troude, V.}, \bibinfo{author}{Lera, S.~C.},
  \bibinfo{author}{Wu, K.} \& \bibinfo{author}{Sornette, D.}
\newblock \bibinfo{title}{Illusions of criticality: Crises without tipping
  points} (\bibinfo{year}{2025}).
\newblock \urlprefix\url{https://arxiv.org/abs/2412.01833}.
\newblock \eprint{2412.01833}.

\bibitem{allesina2012stability}
\bibinfo{author}{Allesina, S.} \& \bibinfo{author}{Tang, S.}
\newblock \bibinfo{title}{Stability criteria for complex ecosystems}.
\newblock \emph{\bibinfo{journal}{Nature}} \textbf{\bibinfo{volume}{483}},
  \bibinfo{pages}{205--208} (\bibinfo{year}{2012}).

\bibitem{sornette2004critical}
\bibinfo{author}{Sornette, D.}
\newblock \emph{\bibinfo{title}{Critical Phenomena in Natural Sciences}}
  (\bibinfo{publisher}{Springer}, \bibinfo{year}{2004}).

\bibitem{troude2025Unifying}
\bibinfo{author}{Troude, V.} \& \bibinfo{author}{Sornette, D.}
\newblock \bibinfo{title}{Unifying framework for amplification mechanisms:
  Spectral criticality, resonance, and nonnormality}.
\newblock \emph{\bibinfo{journal}{Phys. Rev. Res.}}
  \textbf{\bibinfo{volume}{7}}, \bibinfo{pages}{L042048}
  (\bibinfo{year}{2025}).
\newblock \urlprefix\url{https://link.aps.org/doi/10.1103/kdgw-shxf}.

\bibitem{kuramoto1975self}
\bibinfo{author}{Kuramoto, Y.}
\newblock \bibinfo{title}{Self-entrainment of a population of coupled
  non-linear oscillators}.
\newblock \emph{\bibinfo{journal}{International Symposium on Mathematical
  Problems in Theoretical Physics}} \bibinfo{pages}{420--422}
  (\bibinfo{year}{1975}).

\bibitem{acebron2005kuramoto}
\bibinfo{author}{Acebr{\'o}n, J.~A.}, \bibinfo{author}{Bonilla, L.~L.},
  \bibinfo{author}{Vicente, C. J.~P.}, \bibinfo{author}{Ritort, F.} \&
  \bibinfo{author}{Spigler, R.}
\newblock \bibinfo{title}{The {K}uramoto model: A simple paradigm for
  synchronization phenomena} (\bibinfo{year}{2005}).

\bibitem{glass1988clocks}
\bibinfo{author}{Glass, L.} \& \bibinfo{author}{Mackey, M.~C.}
\newblock \emph{\bibinfo{title}{From Clocks to Chaos: The Rhythms of Life}}
  (\bibinfo{publisher}{Princeton University Press},
  \bibinfo{address}{Princeton, NJ}, \bibinfo{year}{1988}).

\bibitem{can-ave97}
\bibinfo{author}{P\'azm\'andi, F.}, \bibinfo{author}{Scalettar, R.} \&
  \bibinfo{author}{Zim\'anyi, G.}
\newblock \bibinfo{title}{Revisiting the theory of finite size scaling in
  disordered systems}.
\newblock \emph{\bibinfo{journal}{Physical Review Letters}}
  \textbf{\bibinfo{volume}{79}}, \bibinfo{pages}{5130--5133}
  (\bibinfo{year}{1997}).

\bibitem{JohSor-can98}
\bibinfo{author}{Johansen, A.} \& \bibinfo{author}{Sornette, D.}
\newblock \bibinfo{title}{Evidence of discrete scale invariance by canonical
  averaging}.
\newblock \emph{\bibinfo{journal}{Int. J. Mod. Phys. C}}
  \textbf{\bibinfo{volume}{9}}, \bibinfo{pages}{433--447}
  (\bibinfo{year}{1998}).

\bibitem{abrams2004chimera}
\bibinfo{author}{Abrams, D.~M.} \& \bibinfo{author}{Strogatz, S.~H.}
\newblock \bibinfo{title}{Chimera states for coupled oscillators}.
\newblock \emph{\bibinfo{journal}{Physical Review Letters}}
  \textbf{\bibinfo{volume}{93}}, \bibinfo{pages}{174102}
  (\bibinfo{year}{2004}).

\bibitem{gomez2011explosive}
\bibinfo{author}{G{\'o}mez-Garde{\~n}es, J.}, \bibinfo{author}{G{\'o}mez, S.},
  \bibinfo{author}{Arenas, A.} \& \bibinfo{author}{Moreno, Y.}
\newblock \bibinfo{title}{Explosive synchronization transitions in scale-free
  networks}.
\newblock \emph{\bibinfo{journal}{Physical Review Letters}}
  \textbf{\bibinfo{volume}{106}}, \bibinfo{pages}{128701}
  (\bibinfo{year}{2011}).

\bibitem{Buzsaki2004}
\bibinfo{author}{Buzs{\'a}ki, G.} \& \bibinfo{author}{Draguhn, A.}
\newblock \bibinfo{title}{Neuronal oscillations in cortical networks}.
\newblock \emph{\bibinfo{journal}{Science}} \textbf{\bibinfo{volume}{304}},
  \bibinfo{pages}{1926--1929} (\bibinfo{year}{2004}).

\bibitem{Buzsaki2006}
\bibinfo{author}{Buzs{\'a}ki, G.}
\newblock \emph{\bibinfo{title}{Rhythms of the Brain}}
  (\bibinfo{publisher}{Oxford University Press}, \bibinfo{year}{2006}).

\bibitem{Buzsaki2012}
\bibinfo{author}{Buzs{\'a}ki, G.} \& \bibinfo{author}{Wang, X.-J.}
\newblock \bibinfo{title}{Mechanisms of gamma oscillations}.
\newblock \emph{\bibinfo{journal}{Annual Review of Neuroscience}}
  \textbf{\bibinfo{volume}{35}}, \bibinfo{pages}{203--225}
  (\bibinfo{year}{2012}).

\bibitem{orellana2023timeseriesMAG}
\bibinfo{author}{Orellana, L.~H.}, \bibinfo{author}{Kr\"uger, K.},
  \bibinfo{author}{Sidhu, C.} \& \bibinfo{author}{Amann, R.}
\newblock \bibinfo{title}{Comparing genomes recovered from time-series
  metagenomes using long- and short-read sequencing technologies}.
\newblock \emph{\bibinfo{journal}{Microbiome}} \textbf{\bibinfo{volume}{11}},
  \bibinfo{pages}{105} (\bibinfo{year}{2023}).

\bibitem{cheng2024genomeresolved}
\bibinfo{author}{Cheng, M.} \emph{et~al.}
\newblock \bibinfo{title}{Deep longitudinal lower respiratory tract microbiome
  profiling reveals genome-resolved functional and evolutionary dynamics in
  critical illness}.
\newblock \emph{\bibinfo{journal}{Nature Communications}}
  \textbf{\bibinfo{volume}{15}}, \bibinfo{pages}{8361} (\bibinfo{year}{2024}).

\bibitem{microbiome2025}
\bibinfo{author}{Kurokawa, R.} \emph{et~al.}
\newblock \bibinfo{title}{High-resolution gut microbiome profiling reveals
  genome-level synchronized dynamics}  (\bibinfo{year}{2025}).

\bibitem{fyodorov2025nonorthogonal}
\bibinfo{author}{Fyodorov, Y.~V.}, \bibinfo{author}{Gudowska-Nowak, E.},
  \bibinfo{author}{Nowak, M.~A.} \& \bibinfo{author}{Tarnowski, W.}
\newblock \bibinfo{title}{Nonorthogonal eigenvectors, fluctuation-dissipation
  relations, and entropy production}.
\newblock \emph{\bibinfo{journal}{Phys. Rev. Lett.}}
  \textbf{\bibinfo{volume}{134}}, \bibinfo{pages}{087102}
  (\bibinfo{year}{2025}).
\newblock
  \urlprefix\url{https://link.aps.org/doi/10.1103/PhysRevLett.134.087102}.

\end{thebibliography}



\clearpage

\pagebreak


{\LARGE\textbf{Supplementary Materials}}

\vspace{0.5cm}


In this Supplementary Material, we derive the results presented in the main manuscript
based on the reduced dynamic given by
\begin{equation}    \label{eq:reduced_bis}
    \dot{\z}(t) = \Gam\z(t) + \et(t),
    \text{ with }
    \Gam = 
    \begin{pmatrix}
        -\alpha & \beta\kappa \\
        \beta\kappa^{-1} & -\alpha
    \end{pmatrix},
\end{equation}
where $\alpha\ge\beta\ge 0$ controls the spectrum
i.e. $-\lambda_{\pm} = -\alpha\pm \beta$ are the eigenvalues;
and $\kappa\ge 1$ is the degree of non-normality 
i.e. $\kappa = 1$ the system is normal, $\kappa> 1$ the system is non-normal.
The noise terms $\et(t)$ is given such that its variance is
\begin{equation}
\mw{\et(t)}{\et(t')^\top} = \B\,\delta(t-t'),
\label{eq:S3_reduced}
\end{equation}
and $\B$ is a positive definite noise covariance matrix, that we write as
\begin{equation}
    \B = \begin{pmatrix}
        \sigma_1^2 & \rho\sigma_1\sigma_2 \\
        \rho\sigma_1\sigma_2 & \sigma_2^2
    \end{pmatrix},
    \label{eq:cov_red2}
\end{equation}
where $\sigma_i ^2$ are the variance of each noise component $\eta_i$,
and $\rho$ their correlation in the non-normal subspace.
In particular, the reduced system contains no oscillatory modes and admits no Hopf bifurcation.

We will also provide additional results supporting the main claim in the main manuscript


\section{Exact power spectrum of the reduced non-normal dynamics}


In this section, we characterize the spectral content of the reduced non-normal dynamics \eqref{eq:reduced}.

The matrix-valued power spectral density of $\z(t)$ is given exactly by the resolvent expression
\begin{equation}
\Ss(f)
= (\Gam - 2\pi if \I)^{-1}\,
\B\,
(\Gam^\top + 2\pi i f \I)^{-1},
\label{eq:S4_Somega}
\end{equation}
where $f$ is a frequency,
which fully determines the spectral properties of the dynamics.

\subsection{Power-spectrum matrix}

We now compute the full power-spectrum matrix associated with the reduced two-dimensional linear dynamics.
Rather than Fourier transforming each entry of the lead--lag covariance matrix $\C(\tau)$ explicitly,
\begin{equation}
    \Ss(\omega) = \int_{-\infty}^{\infty}\C(\tau)\,e^{i\omega\tau}\,d\tau ,
\end{equation}
we take advantage of the standard resolvent representation of the spectral density for linear stochastic systems.
For the reduced dynamics \eqref{eq:reduced}, the power-spectrum matrix can be written in compact form as
\begin{equation} \label{eq:S_resolvent}
    \Ss(\omega)= -\Gam(\omega)^{-1}\B\left(\Gam(\omega)^{\dag}\right)^{-1},
    \qquad
    \Gam(\omega) = \Gam - i\omega\I ,
\end{equation}
where are writting $\omega = 2\pi f$ to simplifies the computation.
This expression highlights that the spectral properties are governed by two distinct ingredients:
the \emph{resolvent} $\Gam(\omega)^{-1}$,
which encodes the linear dynamical response, and the noise covariance $\B$,
which injects fluctuations into the system.

To make the role of non-normality explicit, we now exploit the decomposition,
\begin{equation}
    \Gam=\Sig\,\Gam_0\,\Sig^{-1},~
    \Sig=
    \begin{pmatrix}
        \kappa^{1/2} & 0 \\
        0 & \kappa^{-1/2}
    \end{pmatrix},~
    \Gam_0 =
    \begin{pmatrix}
        -\alpha & \beta \\
        \beta & -\alpha
    \end{pmatrix},
\end{equation}
together with the corresponding decomposition of the noise covariance,
\begin{equation}
    \B=\Sig\,\B'\,\Sig,
    \qquad
    \B'=
    \begin{pmatrix}
        \kappa_\sigma^{-1} & \rho\\
        \rho & \kappa_\sigma
    \end{pmatrix},
    \qquad
    \kappa_\sigma=\frac{\sigma_2}{\sigma_1}\kappa .
\end{equation}
Inserting these relations into \eqref{eq:S_resolvent} yields the similarity form
\begin{equation} \label{eq:S_similarity}
    \Ss(\omega)=
    \Sig\,
    \Ss_0(\omega)\,
    \Sig,
   ~~
    \Ss_0(\omega)=
    -\Gam_0(\omega)^{-1}\B'\left(\Gam_0(\omega)^\dag\right)^{-1},
\end{equation}
with $\Gam_0(\omega)=\Gam_0-i\omega\I$.
Thus, the full spectral structure of the non-normal system is obtained from the ``balanced'' spectrum $\Ss_0(\omega)$,
subsequently dressed by the non-normal scaling $\Sig$.

The inverse resolvent of $\Gam_0(\omega)$ can be computed explicitly,
\begin{equation}
    \Gam_0(\omega)^{-1}
    =-\frac{1}{\left(\lambda_+-i\omega\right)\left(\lambda_- - i\omega\right)}
    \begin{pmatrix}
        \alpha + i\omega & \beta \\
        \beta & \alpha + i\omega
    \end{pmatrix},
\end{equation}
where $\lambda_\pm=\alpha\pm\beta$.
Substituting this expression into \eqref{eq:S_similarity},
we finally obtain the explicit form of the power-spectrum matrix,
\begin{equation}    \label{eq:spec_th}
    \begin{split}
        &\Ss(\omega)
        =
        \begin{pmatrix}
            \sigma S_{1}(\omega;\kappa_\sigma)
            &
            \rho S_2(\omega) \\
            \rho S_2(\omega)^*
            &
            \dfrac{1}{\sigma} S_{1}(\omega;1/\kappa_\sigma)
        \end{pmatrix},
        \qquad
        \sigma=\frac{\sigma_1}{\sigma_2}, \\
        & S_{1}(\omega;\kappa_\rho) = \frac{\omega^2 + \omega_{\kappa_\sigma}^2}{\left(\lambda_+^2 + \omega^2\right)\left(\lambda_-^2 + \omega^2\right)}, \\
        & S_2(\omega) = \frac{\left(\omega - i\omega_+\right)\left(\omega - i\omega_-\right)}{\left(\lambda_+^2 + \omega^2\right)\left(\lambda_-^2 + \omega^2\right)} .
    \end{split}
\end{equation}
Here, we have introduced the characteristic frequency scales
\begin{equation}
    \begin{split}
        &\omega_{\kappa_\sigma}^2
        = \left(\beta\kappa_\sigma\right)^2
        + 2\rho\alpha\beta\kappa_\sigma
        + \alpha^2, \\
        &\omega_\pm
        = \frac{\beta}{\rho}\frac{\kappa_\rho-\kappa_\rho^{-1}}{2}
        \pm
        \sqrt{\left(\alpha + \frac{\beta}{\rho}\frac{\kappa_\rho+\kappa_\rho^{-1}}{2}\right)^2
        - \frac{\beta^2}{\rho^2}} .
    \end{split}
\end{equation}
From $\omega = 2\pi f$, we can obtain the power-spectrum in the frequency domain: $\Ss(f)$.

\subsection{Spectrum along the Reaction Mode}

When the system approaches the pseudo-critical regime,
the dynamics becomes dominated by the reaction mode,
whose power spectrum reads
\begin{equation}
    \begin{split}
        &S_{11}(f) = \sigma \frac{f^2 + f_{\kappa_\sigma}^2}{\left(f^2+f_+^2\right)\left(f^2+f_-^2\right)}, \\
        &\text{with } 2\pi f_{\kappa_\sigma}^2 = (\beta\kappa_\sigma)^2 + 2\rho\alpha\beta\kappa_\sigma + \alpha^2 ,
    \end{split}
\label{eq:S4_S11_def}
\end{equation}
where $2\pi f_\pm=\lambda_\pm$ and the eigenvalues $\lambda_\pm=-\alpha\pm\beta$ of $\Gamma$ \eqref{eq:reduced} are real and negative.
Consequently the poles of $S_{11}(f)$ lie on the imaginary axis,
so the spectrum remains smooth for all real frequencies.

The extrema of $S_{11}(f)$ satisfy
\begin{equation}
\frac{d}{df}S_{11}(f)=0 ,
\label{eq:S4_extremum}
\end{equation}
leading to the candidate frequencies
\begin{equation}
\hat f_\pm^2
=
-f_{\kappa_\sigma}^2
\pm
\sqrt{(f_{\kappa_\sigma}^2-f_+^2)(f_{\kappa_\sigma}^2-f_-^2)} .
\end{equation}
Real extrema therefore exist only when the discriminant is positive and $\hat f_\pm^2>0$.
These conditions restrict the parameter ranges in which a pronounced spectral peak can occur,
and show that the position and prominence of such peaks depend sensitively on the coupling parameters.

Beyond these constraints,
the dominant effect of non-normality is the strong reshaping of the spectrum.
In the pseudo-critical regime, the reaction-mode spectrum develops a pronounced high-frequency damping,
which obeys the scaling
\begin{equation}
S_{11}(f) \approx g(f)=\sigma \frac{f_{\kappa_\sigma}^2}{f^4},
\label{eq:S4_1overf4}
\end{equation}
for $\max\{f_+,f_-\}\ll f^2\ll f_{\kappa_\sigma}^2$,
before crossing over to the high-frequency white-noise regime.
This condition is naturally satisfied once the system enters the pseudo-critical regime,
where $f_{\kappa_\sigma}^2\gg f_+^2$.

The resulting $1/f^4$ tail provides the dominant spectral signature of the dynamics.
It reflects the cumulative effect of non-normal amplification,
which effectively integrates overdamped fluctuations twice along the reaction pathway
\cite{troude2025illusion}.
This mechanism generates strong low-frequency organization while preserving overall stability.

Against this structured spectral background,
finite observation windows naturally reveal interior spectral maxima,
whose position reflects the interplay between the low-frequency amplification and the observational bandwidth.
These peaks therefore capture the effective time scales of the amplified stochastic excursions rather than the eigenfrequencies of intrinsic oscillators.

This analytical characterization clarifies the spectral origin of the coherent frequencies observed in the empirical analyses that follow.
They arise from the interaction between non-normal spectral reshaping and finite-time observation,
which together organize stochastic fluctuations into temporally coherent spectral bands.


\section{Irreversibility and Entropy Production: An Intrinsic Arrow of Time}


The observed intermittent cluster coherence and Kuramoto-like alignment are not merely kinematic effects; they reflect the emergence of an intrinsic arrow of time.
Non-normality generates (i) asymmetric lead-lag correlations,
(ii) circulating probability currents in the reduced subspace,
and (iii) a strictly positive entropy production rate.
These signatures of irreversibility arise despite the absence of intrinsic oscillators,
periodic forcing, or phase-coupling mechanisms.

\subsection{Lead-lag imbalance as an empirical arrow of time}

Consider the reduced linearly stable dynamics described by equation \eqref{eq:reduced}.
We define the stationary lagged covariance matrix
\begin{equation}
\C(\tau) := \lim_{t\to\infty}\mw{\z(t)}{\z(t+\tau)^\top},
\qquad \tau\in\mathbb{R},
\label{eq:S3_Ctau_def}
\end{equation}
and its time-asymmetry (lead--lag) component
\begin{equation}
\Delta\C(\tau) := \frac{1}{2}\Big(\C(\tau)-\C(-\tau)\Big).
\label{eq:S3_DeltaC_def}
\end{equation}
We know that, for $\tau>0$, $\C(\tau) = \Sig e^{\Gam^\top \tau}$
and for $\tau<0$,  $\C(\tau) = e^{\Gam \tau}\Sig$,
where $\Sig$ is the stationary covariance matrices of the process.
So, $\C(-\tau)=\C(\tau)^\top$, and the time-asymmetry component is the antisymmetric component of the matrix i.e.
\begin{equation}
\Delta\C(\tau) := \frac{1}{2}\Big(\C(\tau)-\C(\tau)^\top\Big).
\end{equation}
The scalar lead--lag imbalance is then defined as
\begin{equation}
I(\tau) := \frac{1}{\sqrt{2}}\|\C(\tau)-\C(\tau)^\top\|_F,
\label{eq:S3_Imbalance_def}
\end{equation}
which vanishes if and only if time-reversal symmetry holds at lag $\tau$.

\subsection{Stationary solution and lagged covariance}

From \eqref{eq:reduced}, the stationary solution of the reduced stochastic dynamics reads
\begin{equation}
    \z_t = \int_0^t e^{\Gam (t-s)}\, \et_s\, ds .
\end{equation}
Using stationarity and causality, for $\tau\ge 0$,
we can write the lead-lag covariance matrix \eqref{eq:S3_Ctau_def} as
\begin{equation}
    \C(\tau)=\int_0^\infty \G(t)\,\B\G(t+\tau)\,dt,
\end{equation}
where $\G(t)=e^{\Gam t}$ and $\B$ is the covariance matrix of the noise \eqref{eq:cov_red2} in the reduced non-normal subspace.

Following the decomposition introduced above, we write
\begin{equation}
    \G(t)=\Sig\G_0(t)\,\Sig^{-1},
    \qquad
    \Sig=\diag(\sqrt{\kappa},\,1/\sqrt{\kappa}),
\end{equation}
with
\begin{equation}
    \G_0(t)=e^{-\alpha t}
    \begin{pmatrix}
        \cosh(\beta t) & \sinh(\beta t) \\
        \sinh(\beta t) & \cosh(\beta t)
    \end{pmatrix}.
\end{equation}
Similarly, introducing the rescaled noise covariance
\begin{equation}    \label{eq:B_prime}
    \B'=\Sig^{-1}\B\Sig^{-1}
    =\sigma_1\sigma_2
    \begin{pmatrix}
        \kappa_\sigma^{-1} & \rho \\
        \rho & \kappa_\sigma
    \end{pmatrix},
    \qquad
    \kappa_\sigma=\frac{\sigma_2}{\sigma_1}\kappa,
\end{equation}
the covariance matrix becomes
\begin{equation}
    \C(\tau)=\Sig\left[\int_0^\infty \G_0(t)\,\B'\,\G_0(t+\tau)\,dt\right]\Sig .
\end{equation}

Introducing the auxiliary scalar kernels
\begin{widetext}
\begin{equation}
\begin{split}
    C_{\epsilon_1 \epsilon_2}(\tau) &=
    \frac{e^{-\alpha\tau}}{4}
    \int_0^\infty e^{-2\alpha t}
    \left(e^{\beta(t+\tau)} + \epsilon_1 e^{-\beta(t+\tau)}\right)
    \left(e^{\beta t} + \epsilon_2 e^{-\beta t}\right)\, dt \\
    &=
    \frac{e^{-(\alpha-\beta)\tau}}{8}
    \left(\frac{1}{\alpha-\beta} +\epsilon_2 \frac{1}{\alpha}\right)
    +\epsilon_1
    \frac{e^{-(\alpha+\beta)\tau}}{8}
    \left(\frac{1}{\alpha} +\epsilon_2 \frac{1}{\alpha+\beta}\right),
\end{split}
\end{equation}
\end{widetext}
with $\epsilon_1,\epsilon_2\in\{-1,+1\}$.
The lead--lag covariance matrix can be written compactly as
\begin{equation}
    \begin{split}
        &\C(\tau)=\sigma_1\sigma_2
        \begin{pmatrix}
            \sigma C_1(\tau;\kappa_\sigma)
            &
            C_2(\tau;\kappa_\sigma)
            \\
            C_2(\tau;\kappa_\sigma^{-1})
            &
            \sigma^{-1}C_1(\tau,\kappa_\sigma^{-1})
        \end{pmatrix}, \quad \sigma = \frac{\sigma_1}{\sigma_2}, \\
        & C_1(\tau;\kappa_\sigma) = 
        \kappa_\sigma ^2 C_{--}(\tau) + 2\rho\kappa_\sigma \left(C_{+-}(\tau) + C_{-+}(\tau)\right)
        + C_{++}(\tau) \\
        & C_2(\tau;\kappa_\sigma) = 
        \kappa_\sigma C_{-+}(\tau) + \kappa_\sigma ^{-1} C_{+-}(\tau) + \rho \left(C_{++}(\tau) + C_{--}(\tau)\right) .
    \end{split}
    \label{eq:Ctau_matrix}
\end{equation}
This last expression holds for $\tau>0$, for $-\tau$ the lead--lag covariance is given by $\C(-\tau) = \C(\tau)^\dag$,
and so the mid-distance between the lead--lag covariance at time $\tau$ and $-\tau$ is given by
\begin{equation}    \label{eq:th_imbalance}
    \begin{split}
        &\Delta\C(\tau) := \frac{1}{2}\left[\C(\tau) - \C(\tau)^\dag\right] = K_\sigma\Delta(\tau)
        \begin{pmatrix}
        0 & 1 \\ -1 & 0    
        \end{pmatrix}
        , \\
        &\text{where }
        \Delta(\tau) = C_{-+}(\tau) - C_{+-}(\tau)
        ,\; K_\sigma = \frac{\kappa_\sigma-\kappa_\sigma^{-1}}{2}.
    \end{split}
\end{equation}
The noise correlation $\rho$ enters the full lagged covariance $\C(\tau)$ through the symmetric part of the noise covariance matrix.
As a result, it contributes symmetrically to $C_{+-}(\tau)$ and $C_{-+}(\tau)$.
Consequently
therefore, in the antisymmetric combination $\Delta \C(\tau)$,
all $\rho$-dependent terms cancel exactly,
leaving $\Delta\C(\tau)$ governed solely by the non-normal geometry (through  $K_\sigma$) and independent of $\rho$.

\subsection{Lead-Lag Imbalance \& Non-Normality}

We can define a scalar measure, that we call the lead--lag imbalance,
such that
\begin{equation}
    I(\tau) = \left\|\Delta\C(\tau)\right\|_F,
\end{equation}
where $\|\cdot\|_F$ is the Frobineus norm.
In the non-normal subspace \eqref{eq:reduced},
and considering an input noise covariance matrices as \eqref{eq:cov_red2},
we know from \eqref{eq:th_imbalance},
the explicit expression of the lead--lag imbalance is given by
\begin{equation}
    I(\tau) = \frac{\sigma_1\sigma_2}{2\sqrt{2}\alpha}|K_\sigma|\left|e^{-(\alpha-\beta)\tau} - e^{-(\alpha+\beta)\tau}\right|,    
    \label{eq:ll_imb}
\end{equation}
where $K_\sigma$
is the non-normal index, reshaped to consider the input noise asymmetry $\sigma_1/\sigma_2$.
Therefore, when the system is non-normal, or the noise is anisotropic,
reversibility is broken.
This provides a direct empirical signature of irreversibility:
the dynamics selects a time direction even though it is overdamped and contains no oscillators.
One can note the absence of dependence on $\rho$ in \eqref{eq:ll_imb} due to cancellations in the antisymmetric component of $C(\tau)$ contributing to $I(\tau)$.
It is therefore insensitive to structural changes in the support. 
Non-normal support and non-normal imbalance thus capture complementary aspects of the dynamics.

\begin{figure}
    \centering
    \includegraphics[width=0.4\textwidth]{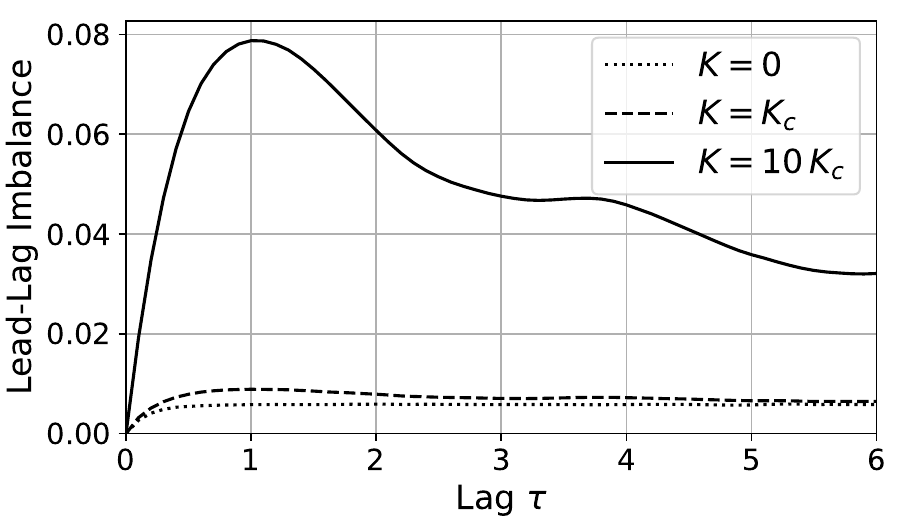}
    \caption{
        \textbf{Emergence of an intrinsic arrow of time from non-normal dynamics.} 
        Lead--lag imbalance $I(\tau)$ \eqref{eq:S3_Imbalance_def} given by (\ref{eq:ll_imb}) as a function of the time lag $\tau$,
        computed from the same synthetic time series shown in the paper,
        for three values of the non-normality index: normal dynamics ($K=0$), pseudo-critical regime ($K=K_c$), and strongly non-normal regime ($K=10K_c$).
    }
    \label{fig:synthetic_imb}
\end{figure}

Figure \ref{fig:synthetic_imb} illustrates this effect by showing the lead--lag imbalance \eqref{eq:S3_Imbalance_def} given by \eqref{eq:ll_imb}, computed over the full system for each time series displayed in the paper. As the non-normality index increases, the imbalance grows, reaches a maximum, and subsequently decays, which is a characteristic signature of non-normal dynamics.
For $K=0$, the imbalance is expected to vanish at all lags. In practice, small nonzero values appear due to statistical bias, yielding a monotonic increase toward a spurious maximum consistent with time-reversal symmetry and equilibrium Ornstein--Uhlenbeck behavior. As $K$ approaches and exceeds $K_c$, however, a pronounced asymmetry emerges over a finite range of lags, revealing a preferred temporal direction despite the absence of intrinsic oscillators or periodic forcing. The amplification and persistence of this lead--lag imbalance provide a direct empirical signature of irreversibility induced by non-normal geometry. This emergent arrow of time is the statistical manifestation of circulating probability currents and foreshadows a strictly positive entropy production rate.

\begin{figure}
    \centering
  \includegraphics[width=0.48\textwidth, trim=20cm 0 0 0, clip]{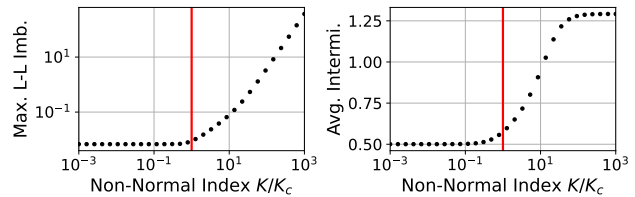}
    \caption{
   \emph{Left panel:} Maximum (over $\tau\in[0,10]$) of the lead-lag imbalance measure \eqref{eq:ll_imb} as a function of $K/K_c$.
\emph{Right panel:} Spectral intermittency measured by the coefficient of variation (CV) \eqref{eq:CV} of the Morlet wavelet power spectrum 
(see figure \ref{fig:synthetic_mwt}) as a function of $K/K_c$.
 }
    \label{fig:phase_transition2}
\end{figure}

Figure~\ref{fig:phase_transition2} shows that temporal irreversibility emerges as a transition as non-normality increases. 
The left panel plots the maximum lead--lag imbalance (\ref{eq:ll_imb}) as a function of the normalized non-normality index $K/K_c$. 
In the normal regime ($K \ll K_c$), the imbalance remains close to zero, indicating time-symmetric fluctuations. 
As $K/K_c$ approaches unity, the imbalance rises sharply, marking the onset of a macroscopic arrow of time associated with circulating probability currents in the reduced non-normal subspace. 
The right panel reports the spectral intermittency measured by the coefficient of variation (CV) (\ref{eq:CV}) of the Morlet wavelet power spectrum
(see figure \ref{fig:synthetic_mwt}).
The CV increases markedly near the same threshold, indicating enhanced temporal variability and intermittent amplification of fluctuations across time scales. 
Together, these observables reveal a coordinated transition near $K/K_c \sim 1$ from a reversible, noise-dominated regime to a pseudo-coherent regime characterized by irreversible dynamics and strongly intermittent spectral activity.

\subsection{Entropy production rate in the reduced non-normal non-equilibrium steady state}

Lead--lag asymmetry is the statistical footprint of circulating probability currents.
In a stationary non-equilibrium steady state (NESS),
the probability current is non-vanishing and forms loops in phase space,
implying the breaking of the detailed balance.

From \eqref{eq:ll_imb}, if the system is non-normal, $\Delta\C(\tau)$ is non-zero,
the process breaks detailed balance and is therefore irreversible.
In this case, the arrow of time measured by $I(\tau)$ is directly associated with the presence of circulating probability currents in the reduced subspace.
Conversely, vanishing probability currents imply detailed balance and hence $\Delta \C(\tau) = 0$.
This establishes a first conceptual bridge: non-normal geometry $\Rightarrow$ lead--lag asymmetry $\Rightarrow$ probability currents.

Irreversibility has a thermodynamic cost.
In stochastic thermodynamics, the entropy production rate $\Phi$ is the canonical quantifier of broken detailed balance and is strictly positive in a NESS with non-vanishing probability currents.

Building on \cite{fyodorov2025nonorthogonal},
the entropy production rate for a linear stochastic system of the form \eqref{eq:reduced} can be expressed as
\begin{equation}    \label{eq:entropy}
    \begin{split}
        &\Phi = \sum_{i,j=\pm} O_{ij}\Lambda_{ij}, \\
        \text{where}\quad
        &O_{ij} = \left(\lv_i\cdot \B\lv_j\right)\left(\rr_j\cdot \B^{-1}\rr_i\right), \\
        &\Lambda_{ij} = -\lambda_i\frac{\lambda_i - \lambda_j}{\lambda_i + \lambda_j}.
    \end{split}
\end{equation}
Here $\rr_i$ and $\lv_i$ denote, respectively,
the right and left eigenvectors associated with the eigenvalue $-\lambda_i$ of the reduced drift matrix in \eqref{eq:reduced},
and $\B$ is the covariance matrix of the noise in the same reduced subspace \eqref{eq:cov_red2}.

In our setting, the reduced matrix has real eigenvalues. As a consequence,
$\Lambda_{ii}=0$ in \eqref{eq:entropy},
and the only non-vanishing contributions to the entropy production rate are the cross terms,
\begin{equation}
    \begin{split}
        &\Lambda_{+-} = -\frac{\beta}{\alpha}\left(\alpha+\beta\right), \\
        &\Lambda_{-+} = \frac{\beta}{\alpha}\left(\alpha-\beta\right).
    \end{split}
\end{equation}
The associated right and left eigenvectors read
\begin{equation}
    \rr_\pm = \frac{1}{\sqrt{\kappa^2+1}}
    \begin{pmatrix}
        \kappa \\ \pm 1
    \end{pmatrix}
    \quad\text{and}\quad
    \lv_\pm =
    \frac{\sqrt{\kappa^2 + 1}}{2\kappa}
    \begin{pmatrix}
        1 \\ \pm \kappa
    \end{pmatrix}.
\end{equation}
It is convenient to work with the rescaled quantity $\kappa_\sigma$ defined in \eqref{eq:B_prime}.
With this notation one finds
\begin{equation}
    O_{\pm\mp} = -\frac{1}{1-\rho^2}\left(\frac{\kappa_\sigma-\kappa_\sigma^{-1}}{2}\right)^2.
\end{equation}
Therefore, the entropy production rate in the reduced non-normal subspace can be written as (see Methods)
\begin{equation}
    \begin{split}
        &\qquad\qquad\Phi = \frac{2\beta^2}{\alpha}\,\frac{K_\sigma^2}{1-\rho^2}, \\
        &K_\sigma := \frac{1}{2}\Big(\kappa_\sigma-\kappa_\sigma^{-1}\Big),
        \qquad
        \kappa_\sigma := \kappa\,\frac{\sigma_2}{\sigma_1}.
    \end{split}
\label{eq:S3_EPR_closed}
\end{equation}
Equation~\eqref{eq:S3_EPR_closed} makes explicit two distinct amplification routes to irreversibility:
geometric amplification via $K_\sigma$ (non-normality) and statistical amplification via $|\rho|\to 1$
(effective noise correlations in the reduced basis).
In particular, $\Phi$ increases sharply as the system enters and crosses the pseudo-critical regime.
Thus, the same mechanism that generates apparent synchronization 
necessarily generates a thermodynamic arrow of time.

\subsection{Entropy-maximizing non-normal support and the extensivity of irreversibility}

The reduced dynamics \eqref{eq:reduced} evolves in a two-dimensional non-normal subspace but is embedded in a high-dimensional system.
A natural and central question is therefore \emph{where} irreversibility resides in the full system and under which conditions it becomes a collective,
macroscopic property.

Let $\PP_N = (\pp_1,\pp_2)$ denote the isometry projecting the full $N$-dimensional system onto the reduced non-normal subspace.
We assume that the stochastic forcing is uncorrelated in the canonical basis,
with heterogeneous variances $v_j$ along each coordinate.
The reduced noise covariance matrix $\B$ is then given by
\begin{equation}
\begin{split}
\sigma_i^2 &= \sum_j p_{i,j}^2\,v_j, \qquad i=1,2,\\
\sigma_1\sigma_2\,\rho &= \sum_j p_{1,j}p_{2,j}\,v_j. 
\end{split}
\label{eq:S3_reduced_noise}
\end{equation}

Equation~\eqref{eq:S3_reduced_noise} makes explicit how effective noise correlations $\rho$ can emerge \emph{purely through projection},
even though the original noise is uncorrelated.
It also immediately clarifies when such correlations are absent.
If the noise is isotropic in the canonical basis ($v_j\equiv v$),
orthogonality of $\pp_1$ and $\pp_2$ implies $\rho=0$.
Likewise, if the two vectors have disjoint supports,
i.e.\ $p_{1,j}p_{2,j}=0$ for all $j$, the cross term vanishes identically.

Conversely, generating large $|\rho|$ requires substantial overlap between the supports of the non-normal mode and its reaction.
A sufficient condition is $|p_{1,j}|=|p_{2,j}|$ across their common support,
which enforces $\sigma_1=\sigma_2$ and distributes noise power symmetrically between the two reduced directions.
Orthogonality then requires that the signs of the products $p_{1,j}p_{2,j}$ vary across coordinates,
inducing a structured partition of the system.

\begin{figure*}
    \centering
    \includegraphics[width=\textwidth]{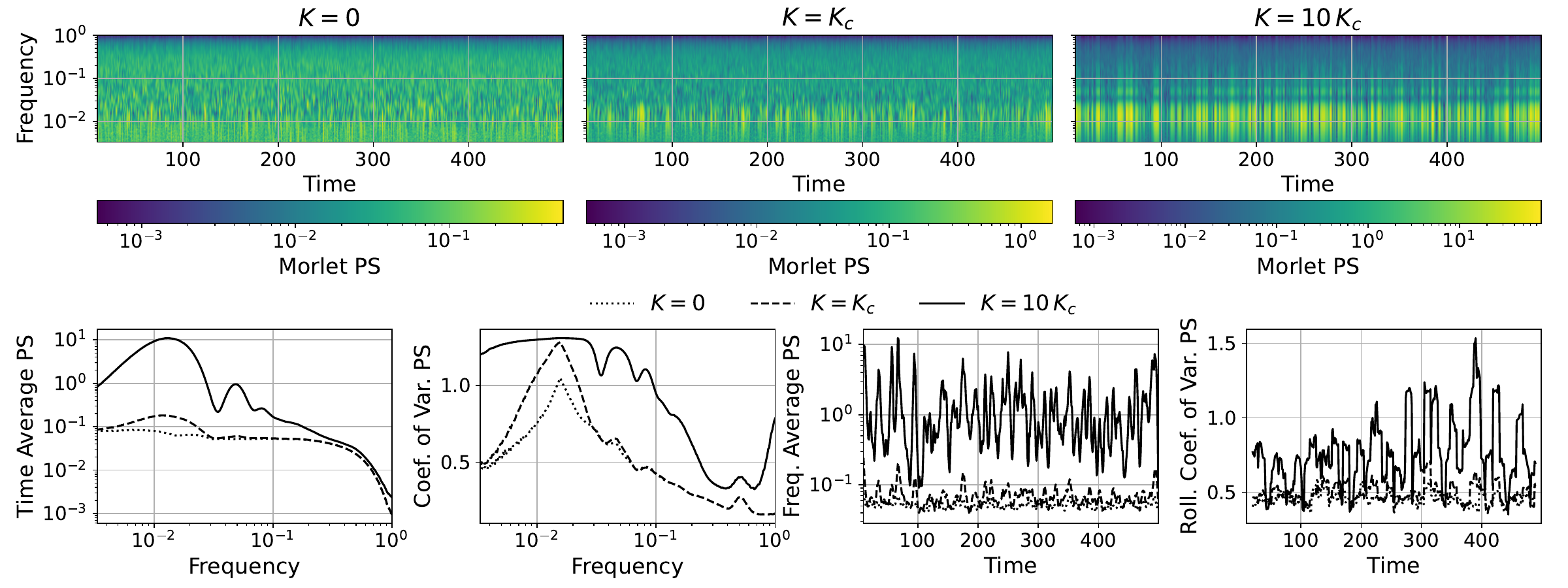}
    \caption{
        \textbf{Morlet wavelet analysis of the synthetic time series in the paper for three values of the non-normality index ($K=0$, $K=K_c$, $K=10K_c$)}.
\emph{Top panels:} Morlet scalograms showing the wavelet amplitude as a function of time and frequency.
\emph{Lower-left panel:} Time-averaged Morlet spectra obtained by averaging the wavelet amplitude over time.
\emph{Lower-center-left panel:} Coefficient of variation (CV) \eqref{eq:CV} of the Morlet amplitude as a function of frequency computed over the full time series.
\emph{Lower-center-right panel:} Smoothed temporal envelope of wavelet activity, obtained from the maximum Morlet amplitude across frequencies at each time and filtered using a Savitzky--Golay filter applied to $\log_{10}$ of this maximum amplitude.
\emph{Lower-right panel:} Coefficient of variation computed over a centered rolling time window and averaged across frequencies, quantifying temporal variability of the wavelet amplitude.
    }
    \label{fig:synthetic_mwt}
\end{figure*}

To make this explicit, we define
\begin{equation}
v_\pm = \sum_{j\,|\,\mathrm{sign}(p_{1,j}p_{2,j})=\pm1} p_{1,j}^2\,v_j,
\label{eq:S3_vpm}
\end{equation}
in terms of which the effective correlation coefficient reads
\begin{equation}
\rho = \frac{v_+ - v_-}{v_+ + v_-}.
\label{eq:S3_rho_vpm}
\end{equation}
This expression shows that $|\rho|$ is maximized when large-variance coordinates predominantly align with one sign of the overlap between $\pp_1$ and $\pp_2$,
while small-variance coordinates align with the opposite sign,
all while preserving orthogonality of the reduced basis.

This geometric mechanism provides a direct maximization principle for the entropy production rate.
Combining \eqref{eq:S3_EPR_closed} with \eqref{eq:S3_rho_vpm},
we see that $\Phi$ is maximized when two conditions are simultaneously met:
\begin{enumerate}
\item the reduced dynamics is strongly non-normal (large $K_\sigma$), amplifying probability currents;
\item the noise variance heterogeneity is redistributed so as to generate strong effective correlations $|\rho|\to 1$ in the reduced subspace.
\end{enumerate}

We quantify how broadly the reaction mode spreads across the system by introducing a \emph{non-normal support} measure. 
We seek a metric that is minimal when the reaction mode is localized on a single component in the canonical basis, $\pp_1=(1,0,\ldots,0)$, and maximal when it is uniformly distributed, $\pp_1=(1,\ldots,1)/\sqrt{N}$. 
For a normalized vector $\pp$, we define
\begin{equation}
s(\pp)=\frac{1}{\sqrt{N}}\sum_{i=1}^{N}|\pp_i|,
\label{eq:support}
\end{equation}
which ranges from $s=N^{-1/2}$ for a fully localized mode to $s=1$ for uniform support. 
Large values $s(\pp)\gg N^{-1/2}$ therefore indicate that the reaction mode has nonzero components along many directions and involves coherent participation across the system.

The macroscopic impact of irreversibility depends directly on this support. 
When $s(\pp_1)\sim N^{-1/2}$, entropy production is effectively localized on a few components and remains weak at the system level. 
By contrast, when $s(\pp_1)=\mathcal{O}(1)$, the entropy production generated in the reduced non-normal subspace spreads coherently across the system and becomes extensive, dominating collective observables such as the lead--lag imbalance and apparent synchronization.

Taken together, these results link non-normal geometry, noise heterogeneity, and thermodynamic irreversibility: entropy production becomes macroscopically observable when strong non-normal amplification coincides with broad reaction-mode support, allowing microscopic fluctuations to generate system-wide circulating currents.

\section{Time-frequency analysis with Morlet wavelet transform}

To further probe the temporal structure of the spectral fluctuations and overcome the limitations of fixed-window Fourier analysis,
we perform a time--frequency decomposition using the Morlet wavelet transform (MWT).
The MWT provides a localized representation of spectral amplitude as a function of time and frequency,
allowing us to distinguish persistent dynamical features from intermittent, estimator-induced effects.

Figure~\ref{fig:synthetic_mwt} displays the Morlet scalograms computed from the same synthetic time series as in the paper, 
averaged across system components.
In the normal regime, wavelet amplitudes remain weak and broadly distributed across frequencies,
with no dominant structure in time, consistent with equilibrium Ornstein--Uhlenbeck fluctuations.
As the system approaches and crosses the pseudo-critical regime,
the Morlet amplitudes become strongly intermittent and 
the spectral energy is  concentrate around a specific frequency band that progressively shifts as a function of time.
This behavior indicates that non-normal amplification promotes episodic concentration of stochastic fluctuations without stabilizing a characteristic oscillatory mode.

Importantly, as the Morlet wavelet transform is not normalized to unit energy at each scale,
the  instantaneous wavelet amplitude at a given frequency quantifies the local strength of fluctuations over the corresponding time scale.
Large temporal variations of this amplitude directly reflect the stochastic amplification and relaxation cycles induced by non-normal dynamics.
In this sense, the MWT reveals how noise-driven fluctuations are intermittently magnified and released over time.
When averaging the Morlet spectrum over time,
one can observe the emergence of local maxima at intermediate frequencies,
whose positions depend on the degree of non-normality.

Additional insights is obtained by examining the coefficient of variation (CV) of the Morlet amplitude at each frequency 
\begin{equation}    \label{eq:CV}
    \text{CV}(f) = \frac{\sqrt{\text{Var}_t\left[P(f)\right]}}{\langle P(f)\rangle_t},
\end{equation}
constructed by averaging over time by using the whole data sets generated when plotted as a function of the frequency,
or over rolling time window and averaged out over the frequency domain when plotted as a function of time.
In the pseudo-critical and strongly non-normal regimes,
the CV is largest at low frequencies and decreases monotonically toward higher frequencies.
Thus, the frequencies carrying the largest average power are also those exhibiting the strongest temporal variability.

Taken together, the Morlet analysis confirms and refines the conclusions drawn from Fourier-based methods.
Non-normal amplification spontaneously organizes stochastic fluctuations into coherent collective episodes characterized by finite, emergent frequencies.
The spectral peaks observed in finite-window Fourier spectra or in time-averaged wavelet representations therefore reflect the spectral signature of pseudo-coherence: transient but highly structured frequency bands generated by non-normal dynamics in a purely overdamped, noise-driven system without intrinsic oscillators.

\end{document}